\documentstyle[pra,aps,preprint,epsf]{revtex}

\begin{document}
\title{Remote Spectral Measurement Using Entangled Photons}
\author{Giuliano Scarcelli, Alejandra Valencia, Samuel Gompers and Yanhua
Shih}
\address{Department of Physics, University of Maryland, Baltimore
County, Baltimore, Maryland 21250} \maketitle\date{17 March, 2003}

\begin{abstract}
By utilizing the frequency anticorrelation of two-photon states
produced via spontaneous parametric down conversion (SPDC), the
working principle of a novel remote spectrometer is demonstrated.
With the help of a local scanning monochromator, the spectral
transmission function of an optical element (or atmosphere) at
remote locations can be characterized for wide range of
wavelengths with expected high resolution.
\end{abstract}

\newpage
Two-photon states generated via SPDC have been a very resourceful
tool for studying fundamental aspects of quantum theory
\cite{EPR}. Recently also their practical applications have been
exploited opening new fields like quantum information processing,
quantum metrology, quantum imaging and quantum lithography
\cite{References}. In this paper we use SPDC as a frequency
anticorrelated two-photon source to demonstrate the working
principle of a novel remote spectrometer: a local scanning
monochromator is located in a laboratory, but it defines the
wavelength measured at remote locations because of the frequency
anticorrelation between photons in a pair emitted by SPDC. The
process is equivalent to carry a ``conjugate monochromator" to
remote locations. The proposed method shows a number of
interesting features: SPDC sources offer the natural possibility
of wide spectral ranges of operation, and due to the frequency
anticorrelation between the two photons in a pair and the
coincidence-like type of detection, it is possible to make the two
detectors operate in very different spectral regions without
affecting the measurement with spurious signals and without
changing the resolution of the measurement determined by the local
monochromator.

The process of SPDC involves passing a pump laser beam through a
nonlinear material, for example, a non-centrosymmetric crystal.
Occasionally, the coherent nonlinear interaction leads to the
annihilation of a high frequency pump photon and the simultaneous
creation of two lower frequency photons, signal and idler, which
satisfy the phase matching conditions \cite{Klyshkobook}:
\begin{equation}
\omega_{p} = \omega_{s} + \omega_{i}, \hspace{5mm} {\bf
k}_{p}={\bf k}_{s} + {\bf k}_{i} \label{eq:phsmtch}
\end{equation}
where $\omega _{j}$, {\bf k$_{j}$ (}j = s, i, p) are frequencies
and wavevectors of the signal (s), idler (i), and pump (p)
respectively.

The schematic setup of the remote spectrometer is shown in
Fig.~\ref{cartoon}.  Photon pairs are generated through the SPDC
process in a local laboratory. The signal photon is sent to a
remote location (e.g. space) passing through the optical element
(or atmosphere) whose transmission spectral function is to be
measured. The idler photon passes through a monochromator in the
laboratory. The signal and the idler are then detected by photon
counting detectors $D_1$, in the space, and $D_2$, in the
laboratory. Each detector is connected to an event timer, an
electronic device that records the registration time history at
which a ``click" detection event on the detector has occurred
\cite{eventimer}. The registration time history of detector $D_1$
of the space station is sent back to the laboratory through a
classical communication channel (telephone, internet etc.). The
two individual registration time-histories are analyzed to achieve
maximum ``coincidences" by shifting the time bases of the two. The
remote spectrometer is now properly set. The spectral function of
the remote spectral filter is obtained by measuring the rate of
coincidence counts at each wavelength defined by the
monochromator.

Perhaps the most important feature of the remote spectrometer is
the enormous range of wavelengths that can be analyzed. This
aspect comes, as previously mentioned, from the frequency
correlation between the signal and idler photons. According to
Eq.~(\ref{eq:phsmtch}), if a pump laser at $400nm$ is used, a
scanning monochromator working in the visible region ($400 nm -
700 nm$) will be able to remotely analyze a virtually infinitely
large range of infrared wavelengths. The resolution of the remote
characterization will be determined by the monochromator's
inherent resolution. Thus, using a high resolution monochromator
in visible wavelengths will permit high resolution calibrations in
infrared wavelengths.

Considering the experimental setup in Fig.~\ref{cartoon}, the
joint detection counting rate, $R_{c}$, of detectors $D_{1}$ and
$D_{2}$, on the time interval $T$, is given by the Glauber formula
\cite{Glauber}:
\begin{equation}\label{coin1}
R_{c}\propto \frac{1}{T}\int_{0}^{T}\int_{0}^{T}\,dt_{1}dt_{2}\,\,
G^{(2)}(t_{1},r_{1}; t_{2},r_{2})
\end{equation}
In Eq.~(\ref{coin1}) $G^{(2)}(t_{1},r_{1}; t_{2},r_{2})$ is the
second order correlation function defined as:
\begin{eqnarray}\label{G2}
G^{(2)}(t_{1},r_{1}; t_{2},r_{2}) & \equiv &
|\left<0\right|E_{2}^{(+)}(t_{2},r_{2})E_{1}^{(+)}(t_{1},r_{1})
\left|\Psi\right>|^{2} =  \mid \psi (t_{1},r_{1}; t_{2},r_{2})\mid
^{2}
\end{eqnarray}
Here  $\psi$ is defined as the effective two-photon wavefunction
\cite{Rubin}. $E_{i}^{(\pm )}(t_{i}, r_{i})$, $i=1,2$, are
positive-frequency and negative-frequency components of the field
at detectors $D_{1}$ and $D_{2}$ that can be written as:
\begin{eqnarray}\label{field}
E^{(+)}_{1}(t_{1},r_{1})=\int\, d\omega \, f(\omega) \,
a(\omega)\, e^{-i[\omega t_{1} - k(\omega)r_{1}]} \hspace*{10mm} \nonumber\\
E^{(+)}_{2}(t_{2},r_{2})=\int\, d\omega \,
\Pi(\omega-\omega_{M})\, a(\omega)\, e^{-i[\omega t_{2} -
k(\omega)r_{2}]}
\end{eqnarray}
where $f(\omega)$ is the spectral function to be measured and
$\Pi(\omega-\omega_{M})$ simulates the spectral function of the
monochromator: a narrow-bandpass function centered at wavelength
$\omega_{M}$.

The signal-idler two-photon state of SPDC can be calculated by
applying the first order perturbation theory of quantum
mechanics\cite{Klyshkobook}. Restricting the calculation to one
dimension and collinear SPDC, the two-photon state is:
\begin{eqnarray}\label{state1}
\left|\Psi\right>= \int_{-\infty}^{\infty}\, d\nu \, \Phi(\nu) \,
a_s^{\dagger}(\omega^{0}_s+\nu) \, a_i^{\dagger}(\omega^{0}_i-\nu)
\, \left|\hbox{0}\right>,
\end{eqnarray}
where $\Phi(\nu)$ is the spectral amplitude of SPDC and is
determined by the wavevector phase matching inside the nonlinear
crystal, $a^{\dagger}$ is the photon creation operator,
$\left|0\right>$ denotes the vacuum state. Here $\omega^{0}_s$ and
$\omega^{0}_i$ are the central frequencies of the signal-idler
radiation field, $\nu$ is a parameter satisfying:
\begin{eqnarray}\label{nu}
\omega_s=\omega^{0}_s +\nu, \,\,\, \omega_i=\omega^{0}_i - \nu,
\,\,\, \omega^{0}_s + \omega^{0}_i = \omega_p,
\end{eqnarray}
Using Eq.~(\ref{field}) and Eq.~(\ref{state1}), and expanding the
wavevector $k$ to the second order in $\nu$, the effective
two-photon wave function becomes:
\begin{eqnarray}\label{6}
\psi(\tau)=\int \, d\nu \, \Phi(\nu) \,f(\omega^{0}_s
+\nu)\Pi(\omega^{0}_i-\nu-\omega_{M}e^{-i\nu\tau} \,
e^{-\frac{i[k''_{1}r_{1}+k''_{2}r_{2}]\nu^{2}}{2}}
\end{eqnarray}
where
$\tau\equiv[(t_{2}-\frac{r_{2}}{u_{2}})-(t_{1}-\frac{r_{1}}{u_{1}})]$;
$u_{1,2}\equiv 1/k'_{1,2}(\omega^{0}_{s,i})$ are the inverse first
order dispersions of the media in which the signal and the idler
propagate. Eq.~(\ref{6}) indicates that $\psi(\tau)$ is the
Fourier transform of a product of four functions.  The second
order dispersion of the media will contribute to the broadening of
the function $G^{(2)}$\cite{Aleja}: depending on the propagation
distance, this broadening may affect the operational decision of
the ``coincidence" time window width. The second order dispersion
can be ``cancelled" by introducing into the optical path a
different type of dispersive material that has the same magnitude
of second order dispersion but an opposite sign\cite{Franson}, see
Eq.~(\ref{6}). Based on these considerations, we will ignore the
second order dispersion in the following. In this case, we can
treat $\Pi(\omega^{0}_i-\nu-\omega_{M})$ as a $\delta$-function if
$f(\omega^{0}_s +\nu)$ and $\Phi(\nu)$ are much wider then
$\Pi(\omega^{0}_i-\nu-\omega_{M})$. $G^{(2)}(\tau)$ becomes:
\begin{eqnarray}\label{7}
G^{(2)}\sim\mid \int \, d\nu \, \Phi(\nu)\,f(\omega^{0}_s
+\nu)\delta(\omega^{0}_i-\nu-\omega_{M}) \, e^{-i\nu\tau} \mid^{2}
\end{eqnarray}
and the coincidence counting rate is then:
\begin{eqnarray}\label{8}
R_{c} \sim |\Phi(\omega^{0}_i -\omega_{M}) \,
f(\omega_{p}-\omega_{M})|^{2}
\end{eqnarray}
Furthermore, if $\Phi(\omega^{0}_i -\omega_{M})$ is relatively
flat compared to function $f(\omega_{p}-\omega_{M})$, which can be
achieved experimentally, Eq.~(\ref{8}) becomes,
\begin{eqnarray}\label{9}
R_{c} \sim |f(\omega_{p}-\omega_{M})|^{2}
\end{eqnarray}
i.e the rate of coincidence counts reproduces exactly the spectral
function of the remote optical element, but reversed in frequency
with respect to the frequency of the pump.

The detailed experimental demonstration setup is shown in the
lower part of Fig.~\ref{cartoon}.  An Argon ion laser line of
$457.9 nm$ was used to pump a 8mm LBO crystal for SPDC.  The LBO
was cut for type II degenerate collinear phase matching.  The LBO
crystal was slightly tilted in the case of non-degenerate
collinear phase matching. After passing through the crystal, the
pump beam was blocked by two mirrors with high reflectivity at the
pump wavelength and by a Newport RG715 color glass filter.  The
orthogonally polarized photon pair was then split by a polarizing
beam splitter. The transmitted signal photons were detected by a
single-photon counting module $D_1$ (Perkin-Elmer SPCM-AQR-14)
after passing through the optical element to be characterized. The
reflected idler photons were sent to a monochromator (CVI Digikrom
CM110) with $2nm$ resolution through a $38mm$ focal length lens. A
$50mm$ focal length lens was placed at its focal distance from the
LBO crystal in order to collect the necessary wide spectrum of
SPDC radiation into the monochromator. The output of the
monochromator was then collected and detected by another
single-photon counting module $D_2$. The photocurrent pulses from
detectors $D_1$ and $D_2$ were then sent to the ``coincidence
counting circuit" with $5ns$ integrating time window.

In order to meet the requirement that led to Eq.~(\ref{9}), in
which we assumed a relatively flat SPDC spectrum,
$\Phi(\omega^{0}_i -\omega_{M})$, compared to the filter function,
$f(\omega_{p}-\omega_{M})$, we needed to collect the entire region
of relevant SPDC spectrum and couple all the wavelengths into the
monochromator with the same efficiency. The choice of the lenses
was made exactly to pursue this objective.

Fig.~\ref{f850}, Fig.~\ref{f885} and Fig.~\ref{f916} report three
typical measurements for bandpass filters centered at $850 nm$,
$885.6 nm$ and $916 nm$ with bandwidths of $10 nm$, $11 nm$ and
$10 nm$, respectively. In the graphs, we provided two scales of
wavelengths, referred to the signal and the idler wavelengths.
These wavelengths can also be read as local ``actually" measured
wavelength ($\lambda$-idler) and ``remote" indirectly measured
wavelength ($\lambda$-signal). The reported single detector
counting rates of $D_2$ are slightly ``tilted" at longer
wavelengths. The tilting slope is mainly determined by the
coupling efficiency of the monochromator \cite{efficiency}. To
account for this, we normalized the coincidence counts accordingly
(see figure captions for details). It is clear from these
experimental data that the remote measurements agree with the
standard laboratory classical spectral transmissivity calibration
curves and with the theoretical predictions.

The authors would like to thank H. Malak, V. Berardi and M.H.
Rubin for helpful discussions and encouragement.  This research
was supported in part by ONR, NSF and NASA-CASPR program.

\begin{figure}[hbt]
\caption{Scheme of a remote spectrometer and the experimental
setup} \label{cartoon}
\end{figure}

\begin{figure}[hbt]
\caption{Experimental characterization of a $10 nm$ bandpass
filter centered at $850 nm$. The solid line is a direct
measurement of the transmissivity function of the $850nm$ spectral
filter by using classical method; hollow squares are the single
counts of detector $D_1$ ($\sim2.5$ Mc/second); filled squares are
the single counts of detector $D_2$ (peak of $\sim10$ Kc/s). The
circles are the normalized coincidence counts weighted by the
single counts of detector $D_2$ (peak of $\sim900$ cc/s).}
\label{f850}
\end{figure}

\begin{figure}[hbt]
\caption{Experimental characterization of a $11 nm$ bandpass
filter centered at $885.6 nm$. The solid line is the standard
characterization; hollow squares are the single counts of $D_1$
($\sim3$ Mc/s); filled squares are the single counts of $D_2$
(peak of $\sim12$ Kc/s). The circles are the normalized cc
weighted by the single counts of $D_2$ (peak of $\sim1100$ cc/s)
.} \label{f885}
\end{figure}

\begin{figure}[hbt]
\caption{Experimental characterization of a $11 nm$ bandpass
filter centered at $916 nm$. The solid line is the standard
characterization; hollow squares are the single counts of $D_1$
($\sim1.5$ Mc/s); filled squares are the single counts of $D_2$
(peak of $\sim10$ Kc/s). The circles are the normalized cc
weighted by the single counts of $D_2$ (peak of $\sim900$ cc/s).}
\label{f916}
\end{figure}

\newpage
\centerline{\epsfxsize=2in \epsffile{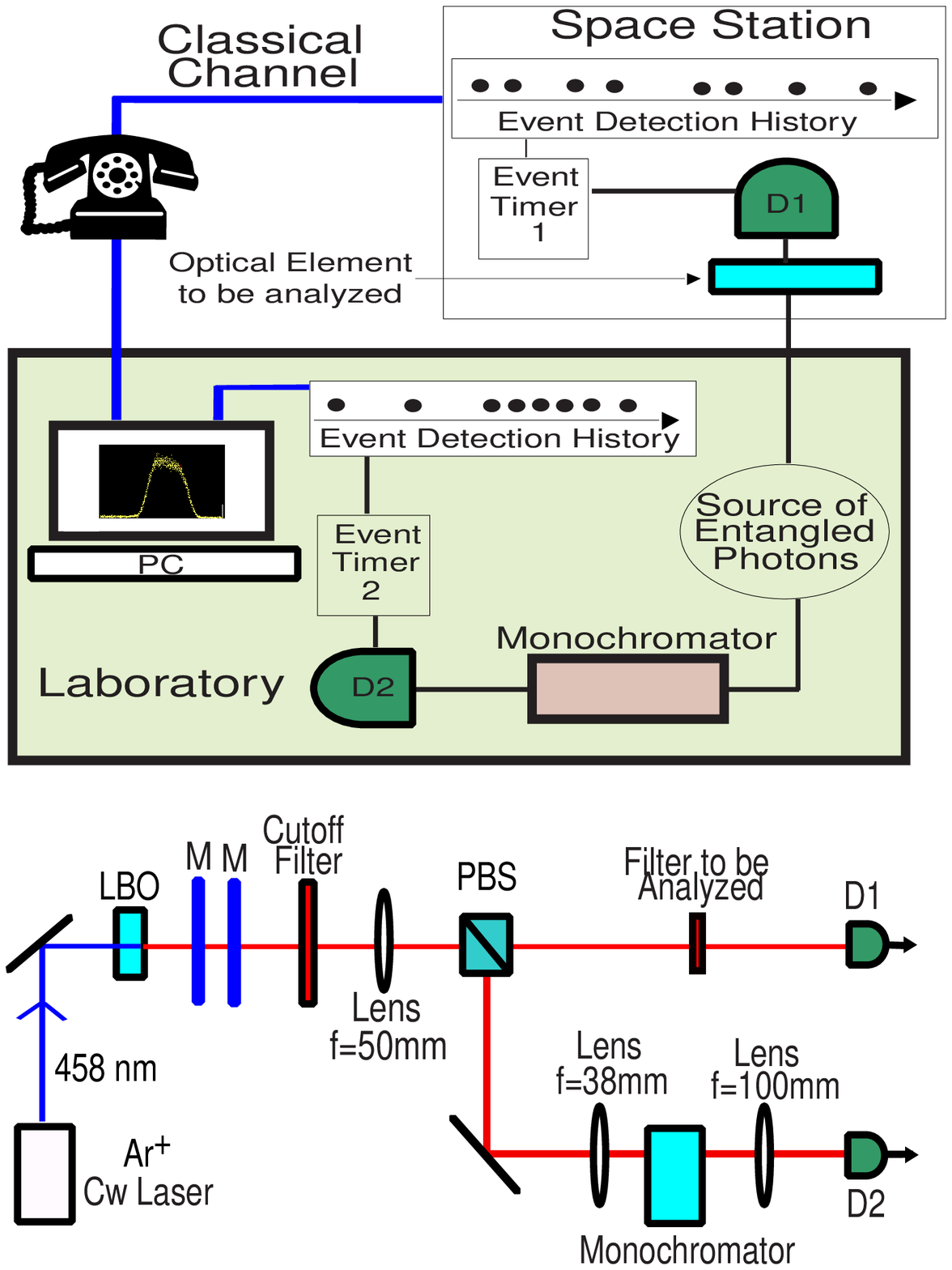}} \vspace{1cm}
Figure \ref{cartoon}.  Giuliano Scarcelli, Alejandra Valencia,
Samuel Gompers , and Yanhua Shih.

\newpage
\centerline{\epsfxsize=2.5in \epsffile{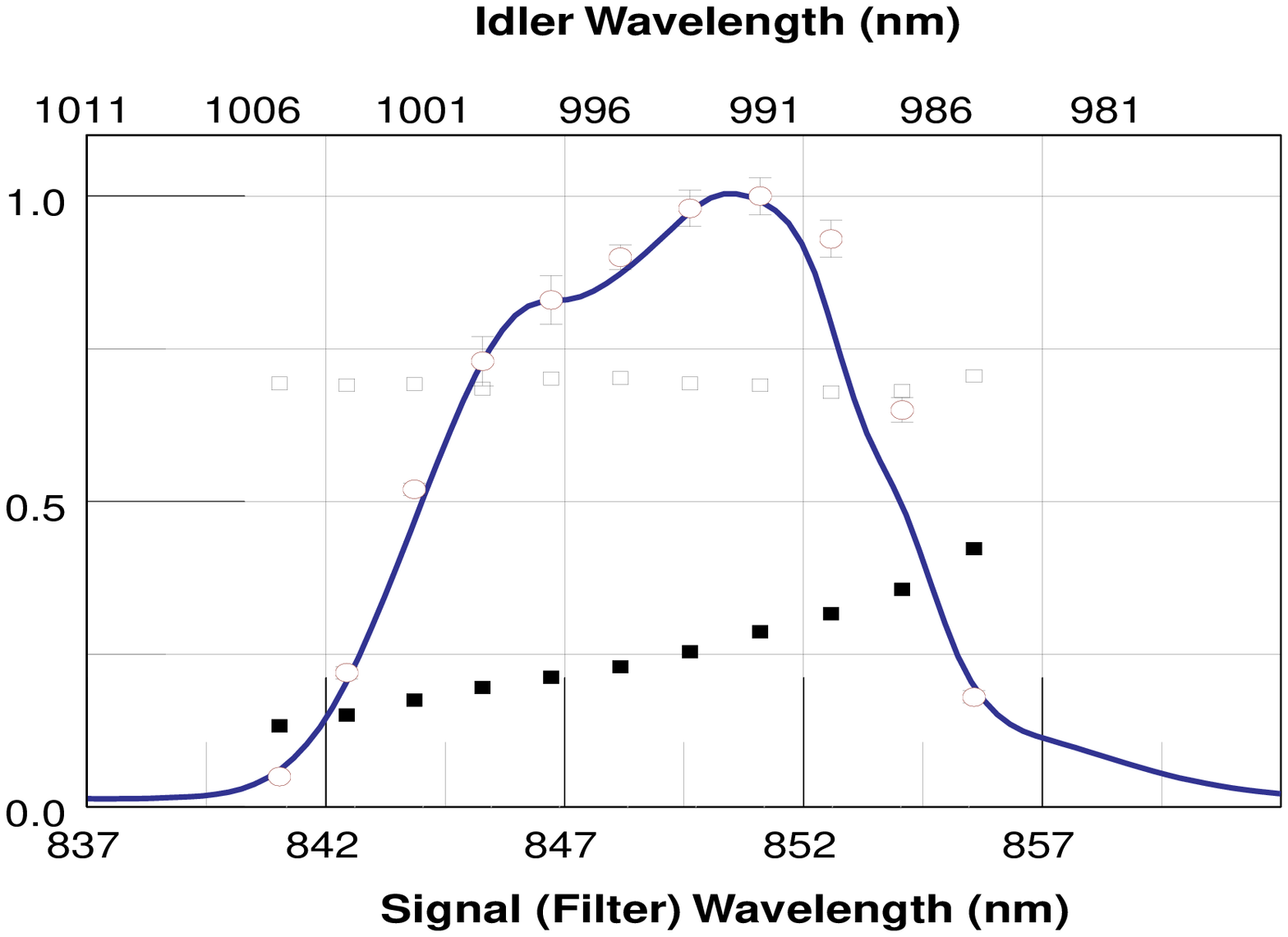}} \vspace{1cm}
Figure \ref{f850}.  Giuliano Scarcelli, Alejandra Valencia, Samuel
Gompers , and Yanhua Shih.

\newpage
\centerline{\epsfxsize=2.5in \epsffile{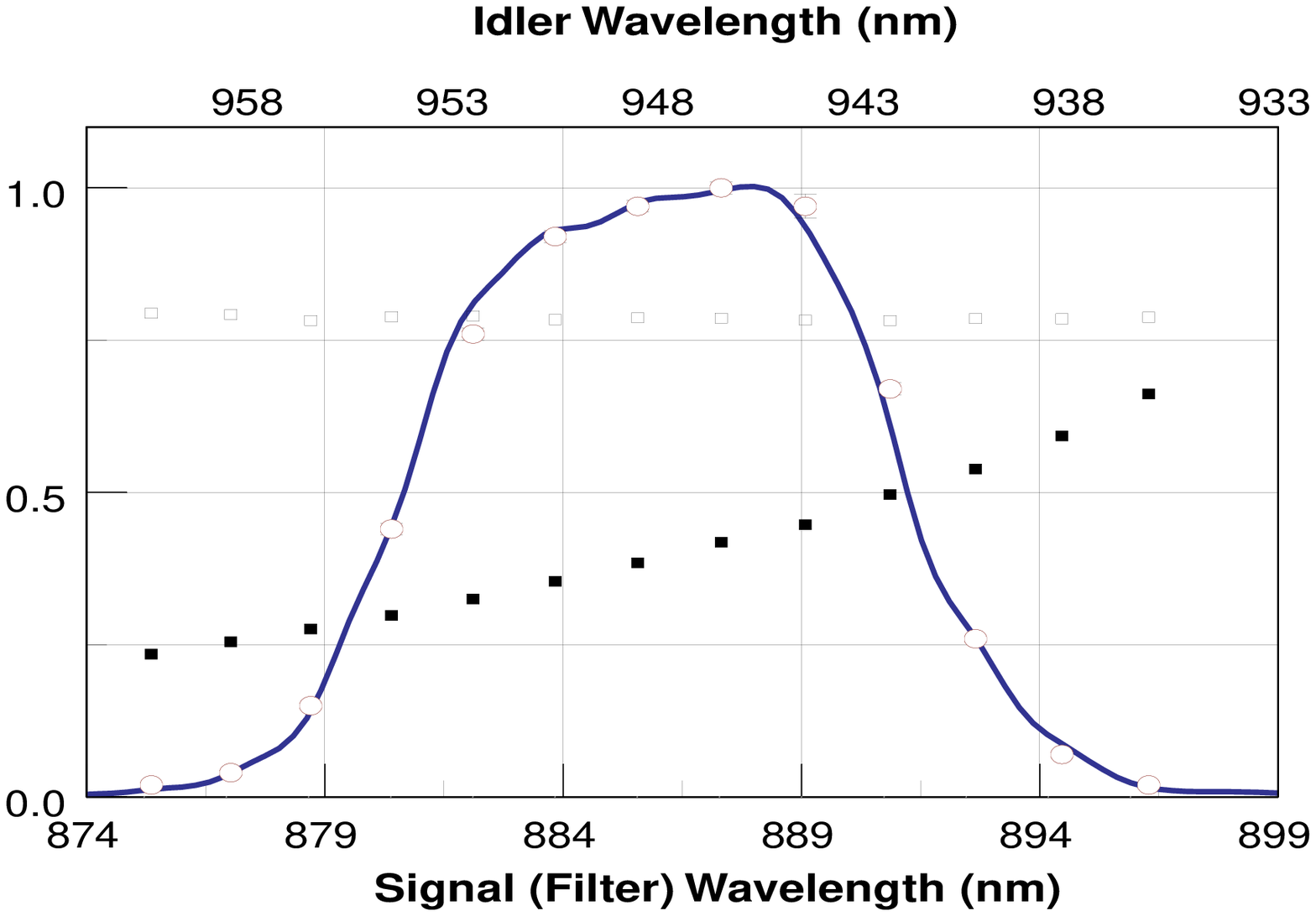}} \vspace{1cm}
Figure \ref{f885}.  Giuliano Scarcelli, Alejandra Valencia, Samuel
Gompers , and Yanhua Shih.

\newpage
\centerline{\epsfxsize=2.5in \epsffile{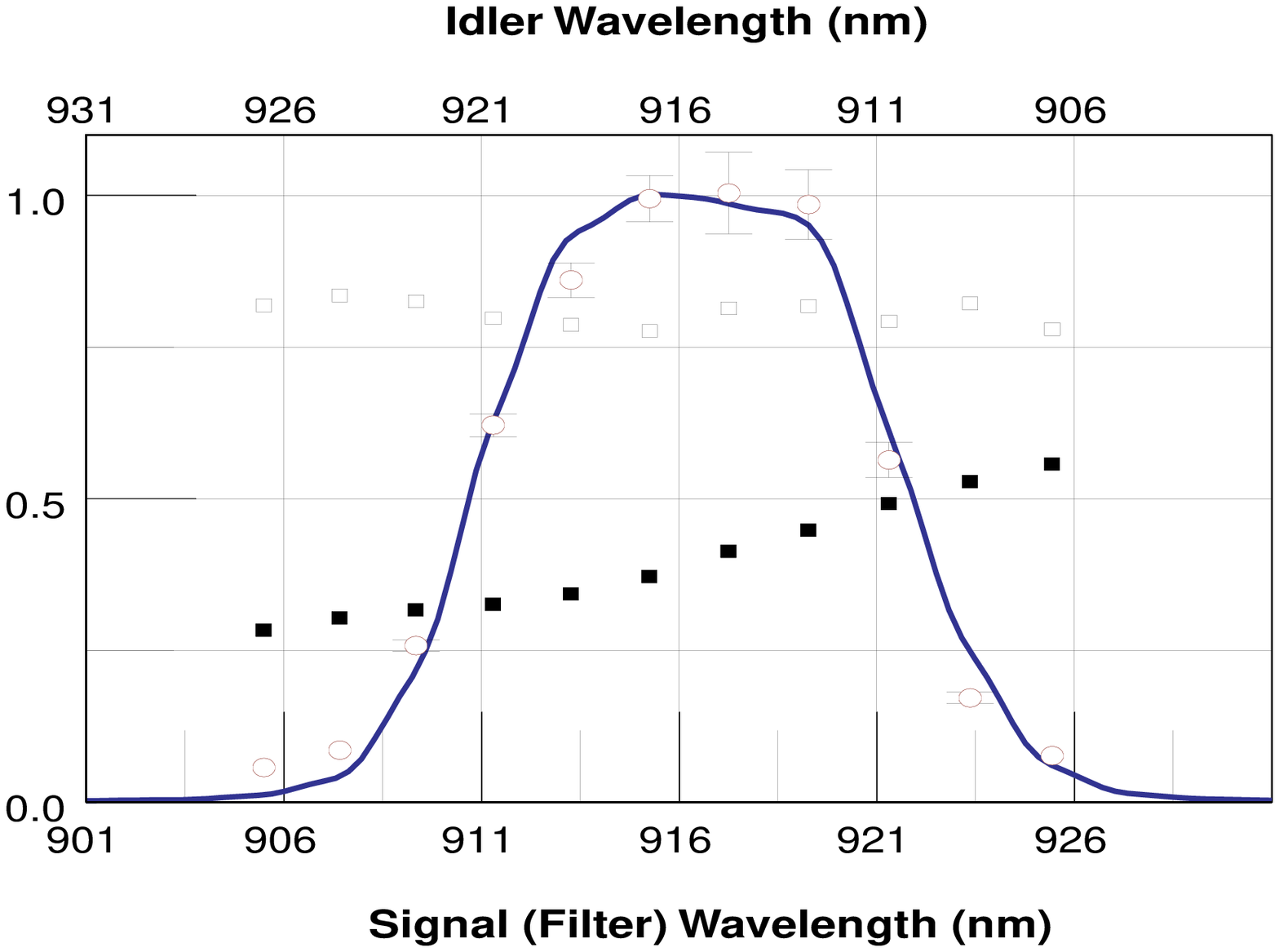}} \vspace{1cm}
Figure \ref{f916}.  Giuliano Scarcelli, Alejandra Valencia, Samuel
Gompers , and Yanhua Shih.

\end{document}